\documentclass[apj]{emulateapj}
\usepackage{epstopdf}
\def\gsim{\ \raise 3pt \hbox{$\rangle$} \kern -8.5pt \raise -2pt \hbox{$\sim$}\ }
\usepackage{graphicx,natbib,color,amsmath,subfigure}
\usepackage{ulem}
\newcommand{\blank}[1]{}
\usepackage{color}

\def\rhessi{{\textit{RHESSI}}}
\def\kw{{Konus-\textit{Wind}}}
\def\mw{{microwave}}

\shorttitle{Coronal thick target flare}
\shortauthors{Fleishman et al.}

\begin{document}


\title{VALIDATION OF THE CORONAL THICK TARGET SOURCE MODEL}

\author{Gregory D. Fleishman\altaffilmark{1}, Yan Xu\altaffilmark{1}, Gelu N. Nita\altaffilmark{1}, \& Dale E. Gary\altaffilmark{1}}
\altaffiltext{1}{Center For Solar-Terrestrial Research, New Jersey Institute of Technology, Newark, NJ 07102}


\begin{abstract}
We present detailed 3D modeling of a dense, coronal thick target X-ray flare using the GX Simulator tool, photospheric magnetic measurements, and microwave imaging and spectroscopy data. The developed model offers a remarkable agreement between the synthesized and observed spectra and images in both X-ray and \mw\ domains, which validates the entire model. The flaring loop parameters are  chosen to reproduce the emission measure, temperature, and the nonthermal electron distribution at low energies {derived} from the X-ray spectral fit, while the remaining parameters, unconstrained by the X-ray data, are selected such as to match the \mw\ images and total power spectra. The modeling suggests that the accelerated electrons are trapped {in} the coronal part of the flaring loop, {but away from} where the magnetic field is minimal, and, {thus,} demonstrates that the data are clearly inconsistent with electron \textit{magnetic} trapping in the weak diffusion regime mediated by the Coulomb collisions. Thus, the modeling supports the interpretation of the coronal thick-target sources as sites of electron acceleration in flares and supplies us with a realistic 3D model with physical parameters of the acceleration region and flaring loop.

\end{abstract}





\section{Introduction}


Coronal flaring loops filled with a relatively dense plasma have been {long} reported based on either soft X-ray (SXR) or \mw\ data  \citep[e.g.,][]{1971SoPh...18..474N, 1980ApJ...240L.111M, 1982ApJ...260..885F}; typically, {high density was seen during the} flare decay phase, when the post-flare loops have become filled with hot plasma as a result of the thermal response on the flare energy release, {thus} no pre-existing dense loops were expected. However, using a favorable combination of microwave and millimeter measurements from a number of radio instruments, \citet{1992ApJ...384..656W} reported an unusually dense impulsive solar burst with ambient plasma density as high as $n_0\sim 5\times10^{11}$~cm$^{-3}$ that in addition was relatively `cold', {i.e. showed no discernable} GOES SXR enhancement above the B3 level. Perhaps, this was the first documented case where the flaring loop was that dense from the very beginning, rather than {becoming dense only during the course of the flare} energy release.

More recently, the unique capability of the Ramaty High Energy Solar Spectroscopic Imager  \citep[\rhessi,][]{2002SoPh..210....3L} to produce X-ray images with high spectral resolution in a broad energy range made it possible to {regularly} detect dense coronal sources by {their} thick-target hard X-ray (HXR) emission \citep{2004ApJ...603L.117V}.  {These \rhessi-detected dense flaring loops are typically hot, but it may be that cold flares as reported by \citet{1992ApJ...384..656W} are completely overlooked without a dedicated radio observing campaign. In the White event, the immeasurably low heating may perhaps not be unexpected due to the weakness of the nonthermal emission,} whose radio flux did not exceed a couple of sfu even at the flare peak time. However, a comparably cold, dense flare {with a} radio flux more than two orders of magnitude larger has been reported by \citet{Bastian_etal_2007} based on the combination of the microwave data jointly obtained with Owens Valley Solar Array \citep[OVSA,][]{ovsa_1984, Gary_Hurford_1994},  Nobeyama Radio Polarimeters \citep[NoRP,][]{Torii_etal_1979}, and the Nobeyama RadioHeliograph \citep[NoRH,][]{Nakajima_etal_1994}. From the detailed analysis of the radio spectral evolution and timing in the event, along with quantitative estimates of the fast electron acceleration efficiency, loss, and energy partitions, \citet{Bastian_etal_2007} concluded in favor of the stochastic acceleration of electrons in this flare, wave-turbulence-mediated transport of the electrons, and accelerated-electron-driven moderate heating of the originally cold, dense plasma of the flaring loop, which was {identified as} the very site where the energy release and electron acceleration happened.

These investigations were limited to relatively rare, non-systematic case studies until \citet{Xu_etal_2008} identified and quantitatively analyzed a distinct class of coronal thick target \rhessi\ flares and derived, from the energy-dependent source sizes, the ranges of likely thermal densities within a few competing models of the source. Similarly to \citet{Bastian_etal_2007}, \citet{Xu_etal_2008} concluded that they deal with the very process of the electron acceleration and estimated such key {acceleration region parameters} as the ambient plasma density and the acceleration region linear size. These data are needed to understand the process of electron acceleration in flares and to develop theoretical models of the acceleration region. A larger number of flares were analyzed by \citet{2012A&A...543A..53G, 2012ApJ...755...32G, 2013ApJ...766...28G} with the same interpretation of the coronal sources as acceleration regions with a high plasma density. Interestingly, if one takes at face value the mean source size and the ambient plasma density derived from the X-ray image analysis of these coronal thick target sources, computes the microwave emission, and compares it with the observed microwave spectrum, a substantial mismatch is {often} found: the low-frequency part of the computed \mw\ spectrum {is predicted to be} strongly suppressed by the Razin-effect, which is not {typically} observed. For example, \citet{2013ApJ...769L..11L} found that to roughly match the model and observed microwave spectrum in one of the coronal thick target flares, one must adopt that the  ambient plasma density is almost one order of magnitude smaller than that derived from the fit of the energy-dependent X-ray source sizes. However, \citet{2014ApJ...787...86J} performed a more accurate analysis of the X-ray data and concluded that the actual source density can in fact be a factor of 5 smaller than that derived from the simplified modeling \citep{Xu_etal_2008, 2012A&A...543A..53G, 2012ApJ...755...32G, 2013ApJ...766...28G}. Taking all these different pieces of the puzzle together leaves us with a large uncertainty about the true physical parameters of the dense coronal thick target sources.

It is well established that considering the X-ray and microwave data together can often improve the flare diagnostics greatly \citep[see, e.g.,][]{Bastian_etal_2007, 2011SSRv..159..225W, Fl_etal_2011,Fl_etal_2013}. To perform a joint study of the X-ray and microwave data on the coronal thick target sources requires a combination of reasonably complementary data sets, which would allow quantitative modeling of both X-ray and microwave emissions and their quantitative comparison with observations. This paper presents one such case from the \citet{Xu_etal_2008} list---the flare of 2002-04-12, which combines the \rhessi\ and \kw\ hard X-ray data and the OVSA \mw\ data with advanced 3D modeling using the recently updated GX Simulator \citep{Nita_etal_2015} and available context observations. We demonstrate that the devised 3D model of the flare offers an excellent fit to the available spectral and imaging X-ray and \mw\ data, which permits a unique determination of the physical parameters in a dense coronal flare source---such as the source size, the number density distribution, the 3D flaring magnetic loop including the range of magnetic field values involved, the 3D spatial distribution of accelerated electrons, and their energy spectrum.

\section{Observations}

\begin{figure}\centering
\includegraphics[width=0.45\textwidth]{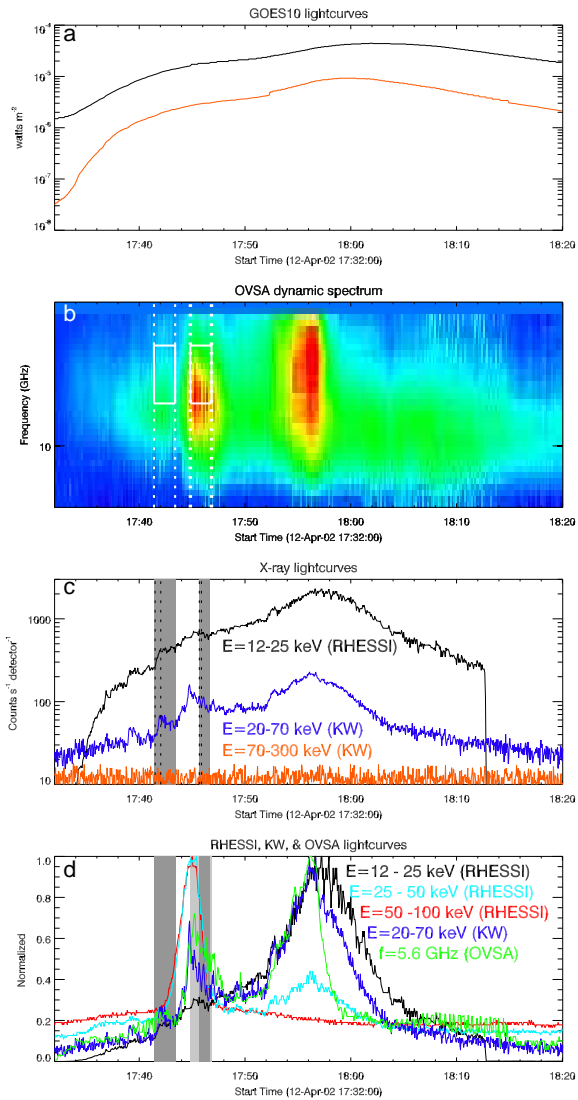}\\
\caption{\label{fig_over} Overview of  April 12, 2002 flare. (a)  GOES (3\hspace{0.1cm}s) lightcurves
as measured by GOES-10 spacecraft. (b) 
OVSA dynamic spectrum. The vertical white dashed lines demarcate the time ranges at the rise (thinner lines) and peak (thicker lines) phases  selected for the OVSA imaging. The white rectangular lines outline the corresponding time-frequency area at the OVSA dynamic spectrum selected for imaging. (c)
\rhessi\ (4 second bins) lightcurves
in one energy range and \kw\ lightcurves in two standard energy ranges (G1 and G2) indicated in the panel. The dark shadow areas show the time intervals selected for \rhessi\ imaging at 12--25~keV range, while the vertical dotted lines demarcate the time ranges selected for the \rhessi\ spectral fit. (d) Normalized \rhessi, \kw, and OVSA light curves. The light shadow shows the time intervals selected for OVSA imaging, while the dark shadow the time intervals selected for \rhessi\ imaging. Note that the light and dark shadows exactly match each other at the first interval. The 50--100~keV light curve does not show emission from the flare; its peak around 17:45~UT is due to charged particle counts during \rhessi\ passage through the radiation belt. A similar particle contribution is seen in the 25--50~keV light curve.
}
\end{figure}

\begin{figure*}[ht] 
\includegraphics[width=0.25\textwidth,clip]{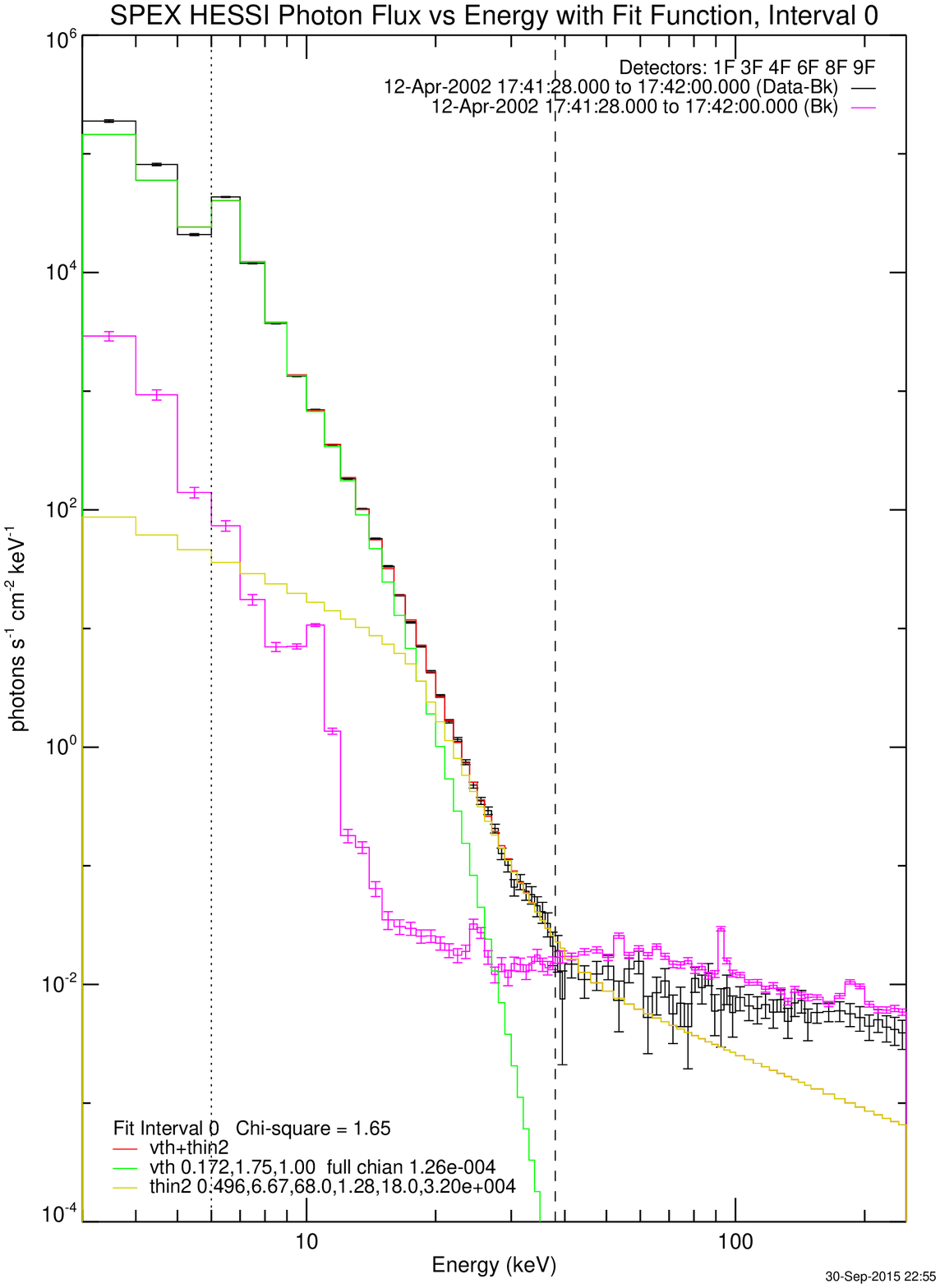}a
\includegraphics[width=0.36\textwidth,clip, bb=49 67 483 488]{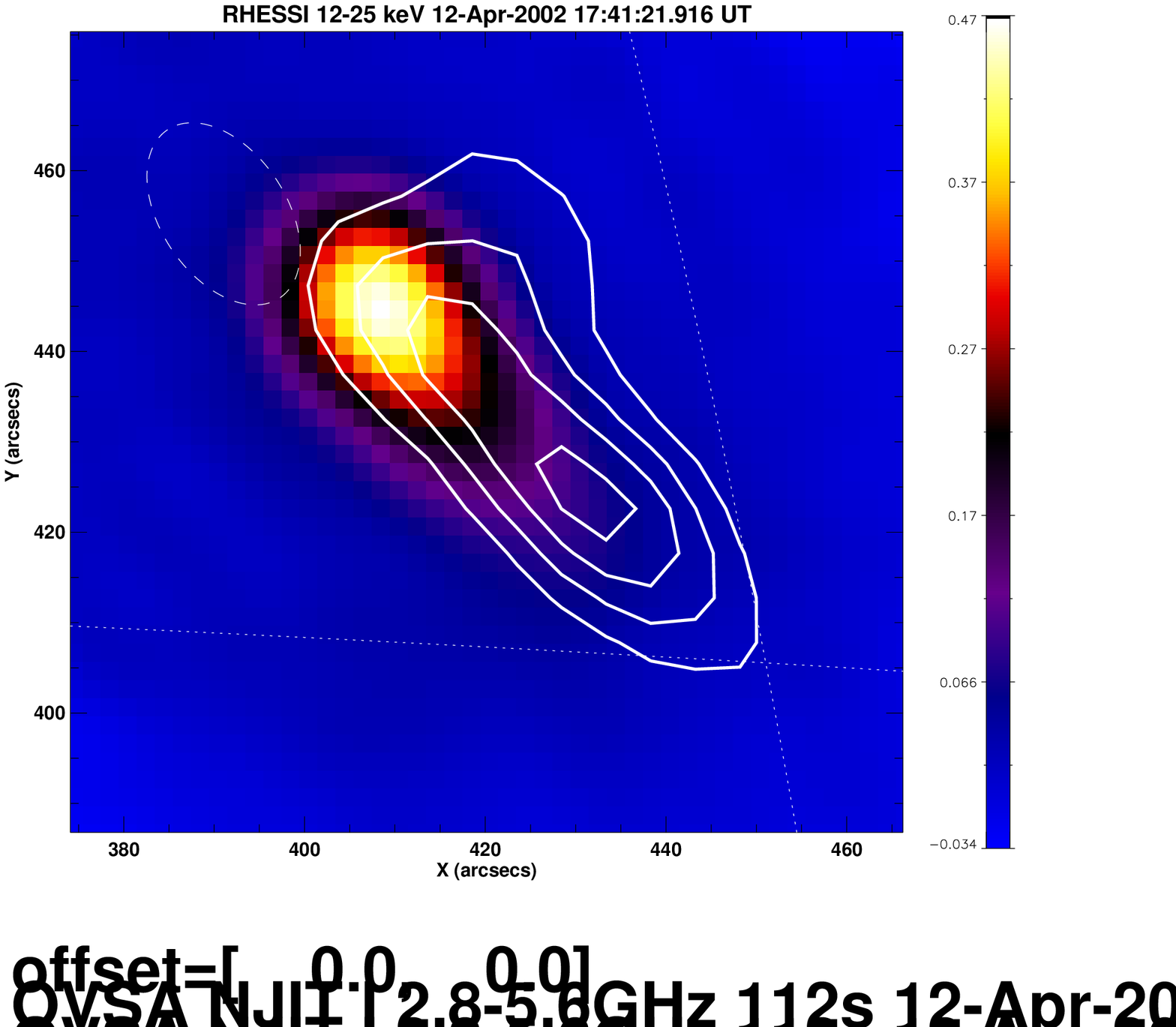}b
\includegraphics[width=0.35\textwidth,clip]{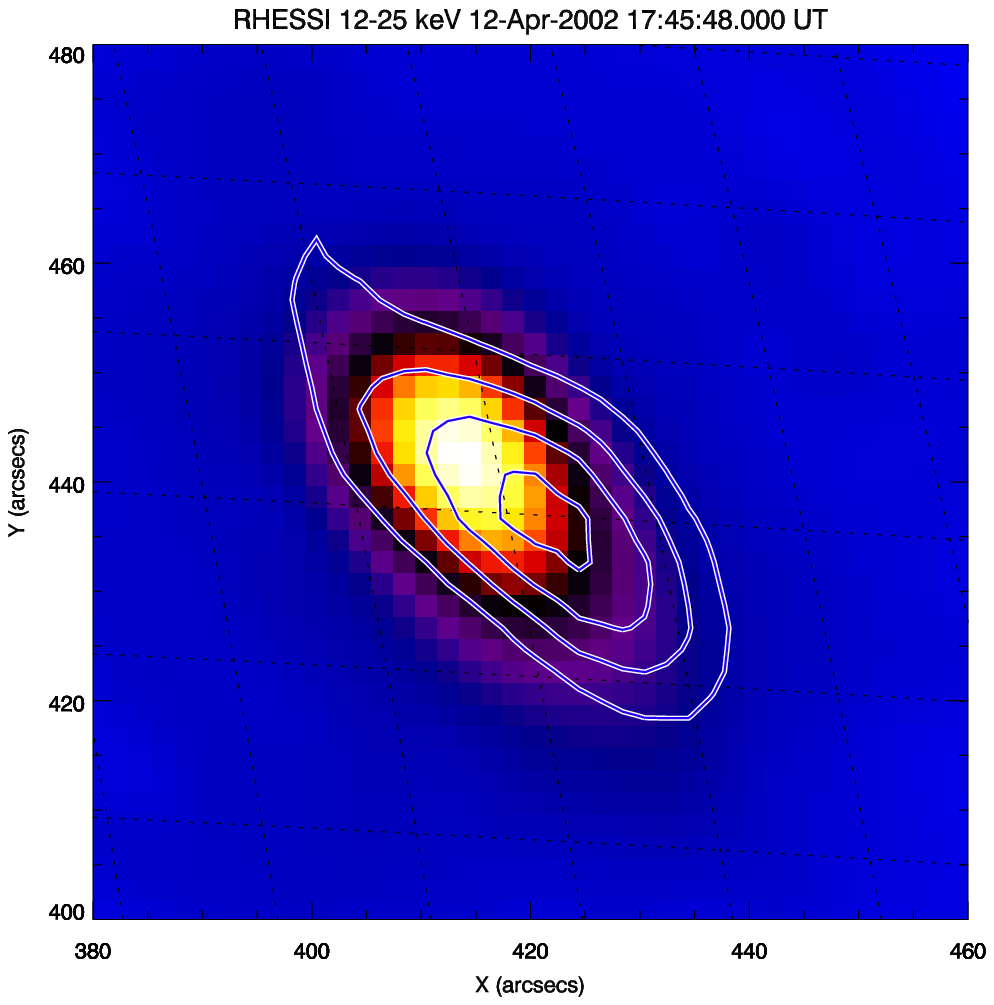}c
\caption{Left: \rhessi\ X-ray spectrum and the thermal plus  power-law thin-target spectral fit for the time range 17:41:28--17:42:00~UT at the flare rise peak. Middle: OVSA time and frequency synthesis 2.8--5.6 GHz  [30,~50,~70,~90]\% contours on top of \rhessi\ 12-25~keV image both taken at the rise phase, 17:41:21.9--17:43:21.9~UT. The dashed oval shows the synthesized OVSA beam. Right: OVSA 5.6 GHz time synthesis (7:44:57.9--17:46:57.9~UT) [30,~50,~70,~90]\%  contours on top of \rhessi\ 12-25~keV image  taken at 17:45:48--17:46:48~UT. An on-line animation of this panel shows the \mw\ image variation as a function of frequency.
\label{fig_images}
}
\end{figure*}
An overview of the solar flare, a GOES class M4.5 flare that occurred on 12 April 2002 at (N22, W25) in AR~9901, is shown in Figure~\ref{fig_over}. This is one of the flares in the list of coronal thick-target flares in the study by \citet{Xu_etal_2008}. HXR data for this event are obtained with \rhessi\ and \kw, and the \mw\ data with OVSA. \rhessi\ and OVSA have complementary spatial, temporal, and spectral/energy resolution, while \kw\ is located in the interplanetary medium, and has a highly stable background \citep{Palshin_etal_2014}, which is helpful {in eliminating complications that can arise} with \rhessi's highly variable background. For example, Figure~\ref{fig_over}d shows that the high-energy \rhessi\ channels are strongly contaminated by the particle contribution from the radiation belt passage during the first peak around 17:45~UT, while the \kw\ light curves are free from such interference. Apparently, there is no burst at the \kw\ channel G2 corresponding to the energy range 70--300~keV. The \rhessi\ 50--100~keV light curve also does not show any enhancement other than the radiation belt particle bump  around 17:45~UT; thus the useful  {X-ray} energy range {available for spectral fit and imaging} is well below 50~keV.
The context GOES-10 data and the line-of-sight magnetogram from the SoHO/MDI \citep{SoHO} are used as well. In particular, the line-of-sight magnetogram is utilized to perform a linear force free field extrapolation needed for the 3D modeling with the GX Simulator \citep{Nita_etal_2015}.

\subsection{HXR imaging and spectra}
\label{S_HXR_imaging_and-spectra}

Hard X-ray (HXR) emission, produced by the energetic particles through bremsstrahlung with the ambient ions, is an important diagnostic of the flare-accelerated electrons in the corona. The key factors that affect characteristics of HXR emission are the distribution of energetic electrons and the number density of the target particles.  Rotating modulation collimators (RMCs) 3-8, yielding a   {nominal FWHM resolution about 7" \citep{2002SoPh..210...61H},   }  were chosen for the image reconstruction using the CLEAN algorithm with pixel size of 2\arcsec.  For comparison, HXR images were generated during  periods of  the corresponding OVSA {data samples}, using time intervals 17:41:22--17:43:22~UT (shown by shadows in Figure~\ref{fig_over}c,d) and 17:45:48 -- 17:46:48~UT\footnote{At the rise phase we used exactly same 2-minutes time interval for both \rhessi\ and OVSA. At the peak time \rhessi\ temporarily changed the attenuator state from 1 to 0, so we used a 1-minute interval for \rhessi\ imaging, that was OK given reasonably high photon statistics at the flare peak phase.} (shown by dark shadows in Figure~\ref{fig_over}c,d), {which are close to the two time intervals selected by \citet{2012ApJ...755...32G} for this event}\footnote{\citet{Xu_etal_2008} studied a single time interval 10 minutes earlier at the very early stage of the flare, where the signal was too low for the analysis performed here.}. Figure~\ref{fig_images}b,c shows HXR images at the standard \rhessi\ range of 12--25~keV, formed mainly by the nonthermal contribution.

In addition to imaging, \rhessi\ provides observations of HXR spectroscopy in photon counts, from which the spectra of electrons {responsible for the emission} can be derived.  We carried out the spectral fitting of integrated HXR spectra {as follows}.  By default, seven {detectors, namely those associated with RMCs}  1, 3, 4, 5, 6, 8, and 9, are summed to generate integrated HXR spectra in photon space. The standard set of 77 nonequal energy bins have been chosen from 3~keV to 250~keV.  A time interval much shorter than that selected for imaging is typically sufficient to produce a meaningful spectral fit. Therefore, it would be reasonable to select a time range for the spectral fit at the very peak of the burst, but the attenuator was temporarily set to ``0'' at the time, {which compromises the spectral data}.   We checked, by analyzing a number of various time intervals {separately for the rise phase and the peak}, that the results are only very weakly sensitive to the exact choice of the time interval for the spectral fitting. Accordingly, we selected the fitting results from two time intervals with attenuator state ``1" for modeling, a 32-second time interval 17:41:28 -- 17:42:00~UT for the rising phase and a 12-second time interval 17:45:48 -- 17:46:00~UT during the peak phase {shown by vertical dotted lines in Fig.~\ref{fig_over}c}.  Since the \rhessi\ data {were} contaminated by the particle event, the spectral fitting is limited to an energy range of 6-38 keV, where such contamination is minimal. {Although the flare belongs to the class of the coronal \textit{thick}-target events, which means that the nonthermal electrons deposit their energy in the coronal part of the loop and do not reach the chromospheric footpoints, here we apply the \textit{thin}-target fitting model because  we are interested in the instantaneous electron distribution at the source, which is the input to the modeling tool GX Simulator.} The fitting results, including both thermal and nonthermal components, are given in Fig.~\ref{fig_images}, left (an example of the rising phase spectrum), and the parameter values are tabulated in Table~\ref{table:1} and Table~\ref{table:2}.

\begin{centering}
\begin{table*}[ht]
\caption{\rhessi\ thermal plus {nonthermal thin-target}  power law spectral fit parameters. 
}
\begin{tabular}{l l l l}
\hline\hline
Parameter & Symbol, units & Rise Phase & Peak Phase\\
Time range & t, UT & 17:41:28--17:42:00 & 17:45:48--17:46:00\\ [0.5ex]
\hline
{\textit{Thermal}:} &  & \\
\qquad Emission Measure & $EM$, cm$^{-3}$ & $0.17\times10^{49}$ & $0.25\times10^{49}$\\
\qquad Temperature      & $T$, MK  & 20 & 21 \\
\textit{Nonthermal:} &  &  \\
\qquad Electron Flux & $F_e\cdot(n_0V)$, el's~cm$^{-2}$~s$^{-1}$ & $0.50\times10^{55}$ & $0.95\times10^{55}$ \\
\qquad Low-energy Cutoff & $E_0$, keV & 18  & 18 \\
\qquad Energy range for fitting & $\Delta E$, keV  &  6--38 & 6--38 \\
\qquad Low-energy Index & $\delta_l$ & $6.7\pm0.5$ & $5.7\pm0.5$ \\
[1ex]
\hline

\end{tabular}

\label{table:1}
\end{table*}
\end{centering}





\begin{figure*} 
\includegraphics[width=0.95\textwidth,clip]{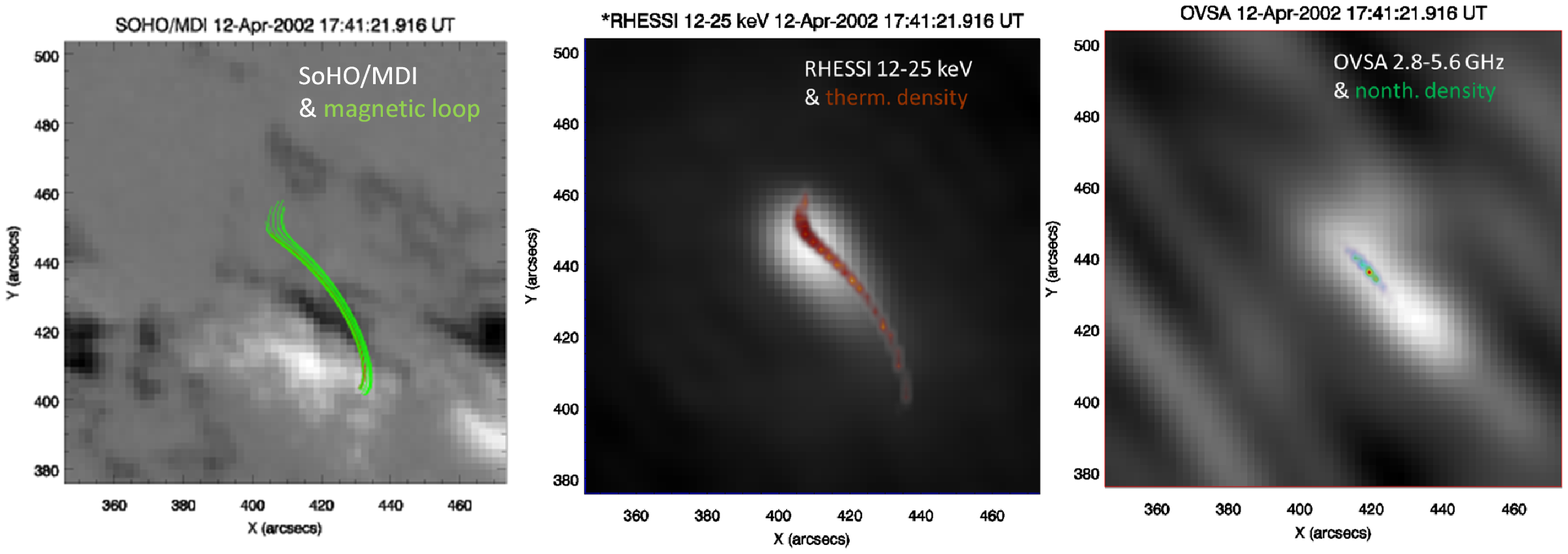}\\
 \hspace{10in}\qquad\qquad a\qquad\qquad \quad\qquad  \qquad\qquad \quad\qquad \qquad\qquad\ \qquad b \qquad\qquad \qquad \qquad\qquad \quad\qquad \qquad\qquad\ c\\
\\
\centerline{\includegraphics[width=0.41\textwidth,clip]{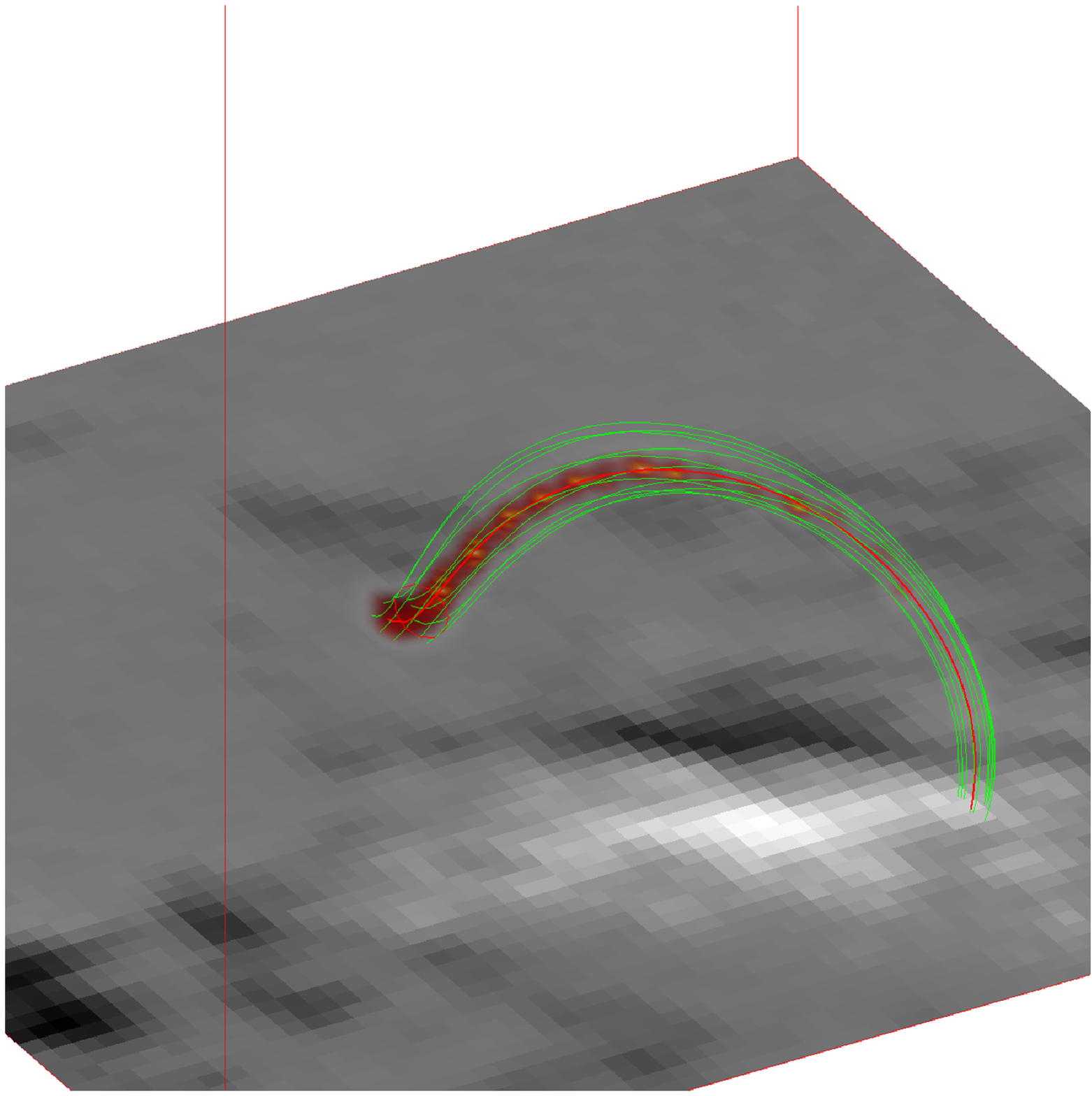}\qquad\qquad 
\includegraphics[width=0.39\textwidth,clip]{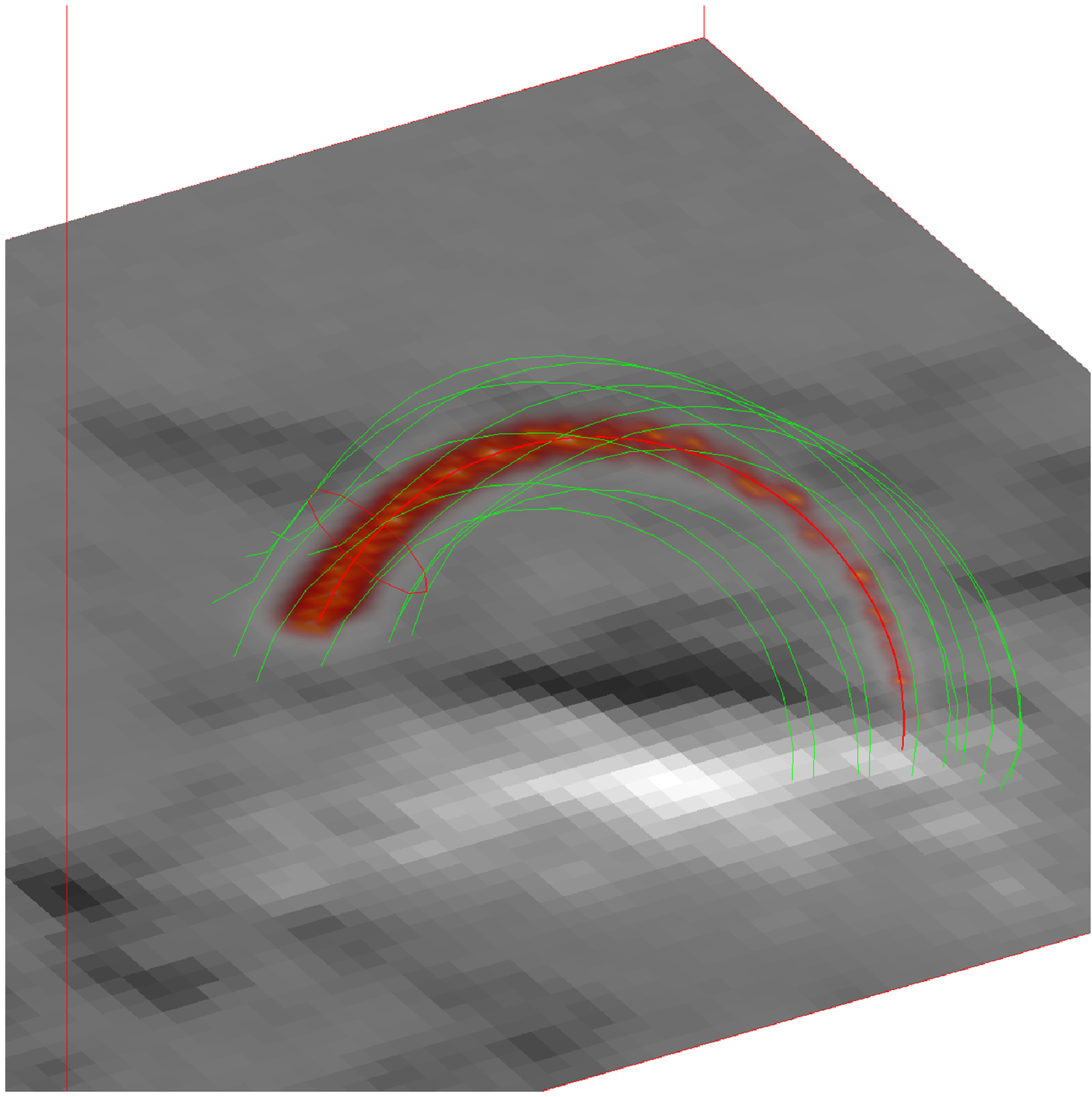}}
d\qquad\qquad \quad\qquad\quad\qquad \qquad \qquad \qquad\qquad\ \qquad\qquad \quad\qquad \qquad\qquad\ \qquad e
\caption{3D model{s} adopted to reproduce HXR and \mw\ spectra and images {for the rise phase (panels a--d) and the peak phase (panel e)}. a: selected magnetic flux tube visualized by a number of field lines (green) surrounding the central field line (red) on top of the  {MDI LOS magnetogram}; b: the {thermal density distribution adopted within this}  magnetic flux tube on top of {the \rhessi\ image (same as in Fig.~\ref{fig_images}); c: nonthermal electron density on top of} the OVSA~2.8--5.6~GHz image (same as in Fig.~\ref{fig_images}); d: perspective view of the model flux tube visualized by the field lines (red and green) and 3D distribution of the ambient plasma density (dark red volume around the central field line); red circle shows a transverse transection of the central field line that demarcates the origin of the coordinate $s$ along the central field line; e: {a similar perspective view of the model built for the flare peak time.}
\label{fig_model_rise}
}
\end{figure*}

\subsection{OVSA imaging}

The spatial distribution of the gyrosynchrotron emission is specified by a combination of the nonthermal electron distribution and the magnetic field distribution along the source. For this event, we employed OVSA imaging at the frequency range 2.8--5.6~GHz around the spectral peak frequency at the rise (17:41:22--17:43:22~UT) and peak time (17:44:58--17:46:58~UT) of the burst, see Fig.~\ref{fig_images}.  The OVSA images at individual frequencies from this range show overall agreement with the frequency synthesis image, but with the lower, frequency-dependent spatial resolution.  The OVSA images are generated using the CLEAN+SELFCAL method from the calibrated OVSA legacy data base, which has recently become available along with the updated OVSA imaging software \citep{2014AAS...22421845N}. The OVSA images at the peak time are overall similar to {those at the rise phase}.



Comparison of the \rhessi\ and OVSA images shows that they have comparable {apparent source} sizes, but are clearly displaced relative to each other as often seen in observations and modeling \citep[see, e.g.,][]{2015SoPh..290...79K, Nita_etal_2015}. The reason for this displacement is generally understood: even {though the} energy-independent spatial distribution of accelerated electrons producing both HXR and \mw\ emissions {is the same}, the HXR emission is weighted with the ambient plasma density, while the \mw\ emission is {weighted by} magnetic field {strength}.

\vspace{-0.25cm}
\section{3D modeling}

In this section we {describe the use of} GX Simulator \citep{Nita_etal_2015} to build a realistic 3D model that would produce both X-ray and \mw\ emissions consistent with available observations {specifically for the two time intervals, the rise phase and the peak, presented above}. According to \citet{Nita_etal_2015} the first step of this modeling is to select the photospheric magnetogram to perform the extrapolation of the magnetic field to the coronal volume. In our case no vector magnetogram is available, so a nonlinear force-free field ($\mathbf{\nabla}\times \mathbf{B} = \alpha(\mathbf{r})\mathbf{B}$, NLFFF) extrapolation could not be performed. Instead, we used the SoHO/MDI line-of-sight magnetogram, which allows either potential or linear force-free field  ($\mathbf{\nabla}\times \mathbf{B} = \alpha\mathbf{B}$ with a constant $\alpha$, LFFF) extrapolation. Although LFFF is known to be globally incorrect \citep[e.g.,][]{Fl_Topt_2013_CED}, it nevertheless can offer an approximately valid local solution needed to approximate a given coronal loop  \citep[e.g.,][]{Nita_etal_2015}, if the constant $\alpha$ parameter is selected in such a way that the implied magnetic connectivity can be established. In particular, we use the \mw\ and {X-ray}  images as guides {to the flaring flux tube location and orientation} and look for a corresponding field line in a magnetic data cube obtained for a given $\alpha$. Adjusting $\alpha$ changes the twist of the magnetic field lines, and, correspondingly, the connectivity in such a way that only a narrow range of $\alpha$ values allows drawing a field line through both HXR and \mw\ sources. In our case, by trial and error, we found that a LFFF model with  $\alpha\sim(6-7)\cdot10^{-10}$~cm$^{-1}$ {provides an appropriate} central field line, whose projection onto the photosphere is elongated in the same direction as both X-ray and \mw\ sources and crosses their centroids. It is important to keep in mind, however, that the LFFF model remains a relatively rough approximation, so no ideally perfect match between the model and data {should} be expected.

\subsection{3D modeling: the rise phase}

For the rise phase we found a best matching {central field line (as shown in Fig.~\ref{fig_model_rise})} in the extrapolated data cube with {nominal value} $\alpha\approx6.6\cdot10^{-10}$~cm$^{-1}$. {Note that because of limitations of the extrapolation method used, the actual $\alpha$ value along the field line (to be compared with that at the peak phase) is slightly different from the nominal one; in our case it is $\alpha\approx(6.4\pm2.0)\cdot10^{-10}$~cm$^{-1}$}. {GX Simulator creates a loop with a transverse radius\footnote{In a general case, an elliptical cross-section with semiaxes $a$ and $b$ can be created.} $a$ 
that varies along the central field line according to the magnetic flux conservation requirement. This radius is then used as a reference scale in the normalized transverse distributions of the thermal and nonthermal particles so a change of $a$ results in consistent changes in both transverse thermal and nonthermal distributions.} Once the model flaring loop has been chosen, the next step is to populate it with the thermal plasma and nonthermal electron distributions. 

To do so, we start with the parameters obtained from the X-ray spectral fit. {The} temperature and total emission measure of the thermal plasma are set to the values in Table~\ref{table:1}: $T   = 20$~MK; $EM = 0.17\times 10^{49}$~cm$^{-3}$. A single value of the temperature within the flaring loop was adopted as $T = 20$~MK to exactly match that from the {HXR spectral} fit. In contrast, the plasma number density is allowed to vary both along and transverse to the flaring loop, Eq.~(1) from \citet{Nita_etal_2015}.  The {normalized} transverse extent of the density distribution {is taken to be} a default 2D Gaussian distribution, Eq.~(1) from \citet{Nita_etal_2015}.  The number density at the central field line location was chosen to be $n_0=2\cdot10^{11}$~cm$^{-3}$, which, being combined with the model flux tube volume {iteratively} determined below, ensures that the integrated emission measure is {close to} $EM = 0.17\times 10^{49}$~cm$^{-3}$ as derived from the X-ray spectral fit{, while the total number of the thermal electrons in the flaring loop volume is $n_0 V=2.25\times10^{37}$}. Note that this $n_0$ gives the density at the central field line (with only a minor variation along the field line following the hydrostatic atmosphere with the temperature $T = 20$~MK), while the off-center density decreases according to the {adopted} gaussian distribution. The top view of the spatial distribution of the ambient plasma density, {superposed} the \rhessi\ image, is given in Fig.~\ref{fig_model_rise}.

The shape of the spatial distribution of the nonthermal electrons in the \textit{transverse} direction is selected to be the same as that of the thermal electrons within the zero-order assumption that both electron acceleration and plasma heating are driven by the same, or closely related, energy release process(es). Eventually, the source area is selected {such that} the model \mw\ low-frequency spectrum matches the observed one. Indeed, the flux level of the optically thick emission is specified by a product of the effective energy of the particles radiating at the given frequency and the source area. This effective energy is specified by the electron energy spectrum {(obtained from the \rhessi\ HXR spectrum)} and the magnetic field value {(obtained from our selection of core field line)}. Thus, the low-frequency \mw\ spectrum yields {a unique value for the} source area. The nonthermal electron distribution \textit{along} the flux tube was initially selected to have a maximum around the geometrical top of the loop and to extend over the spatial locations of both X-ray and \mw\ sources. Ultimately, the shape and central location of this distribution were adjusted {iteratively} by generating a series of X-ray and \mw\ images and {comparing} with the observed ones.  The top view of the {eventually} adopted distribution is shown in Fig.~\ref{fig_model_rise} on top of the OVSA frequency synthesis image {covering} 2.8--5.6~GHz. 

\begin{figure*} 
\includegraphics[width=0.31\textwidth,clip]{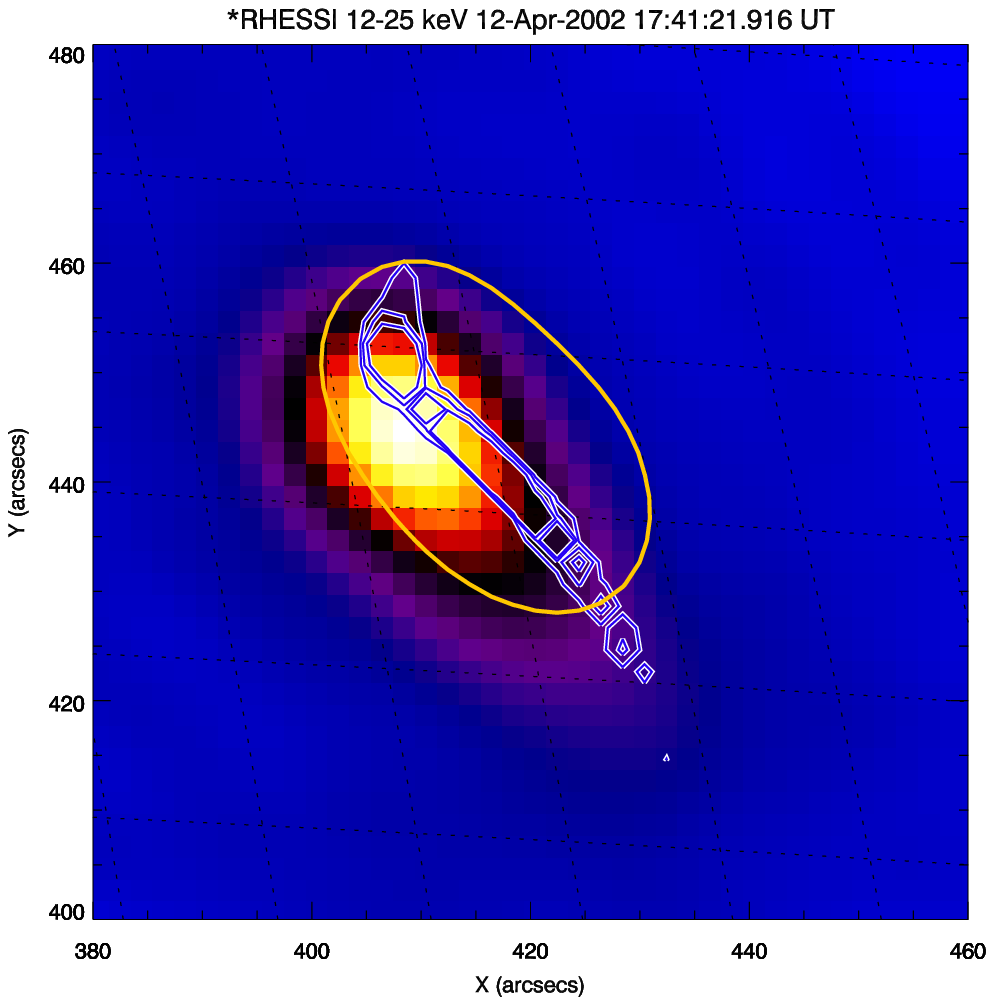}a
\includegraphics[width=0.31\textwidth,clip]{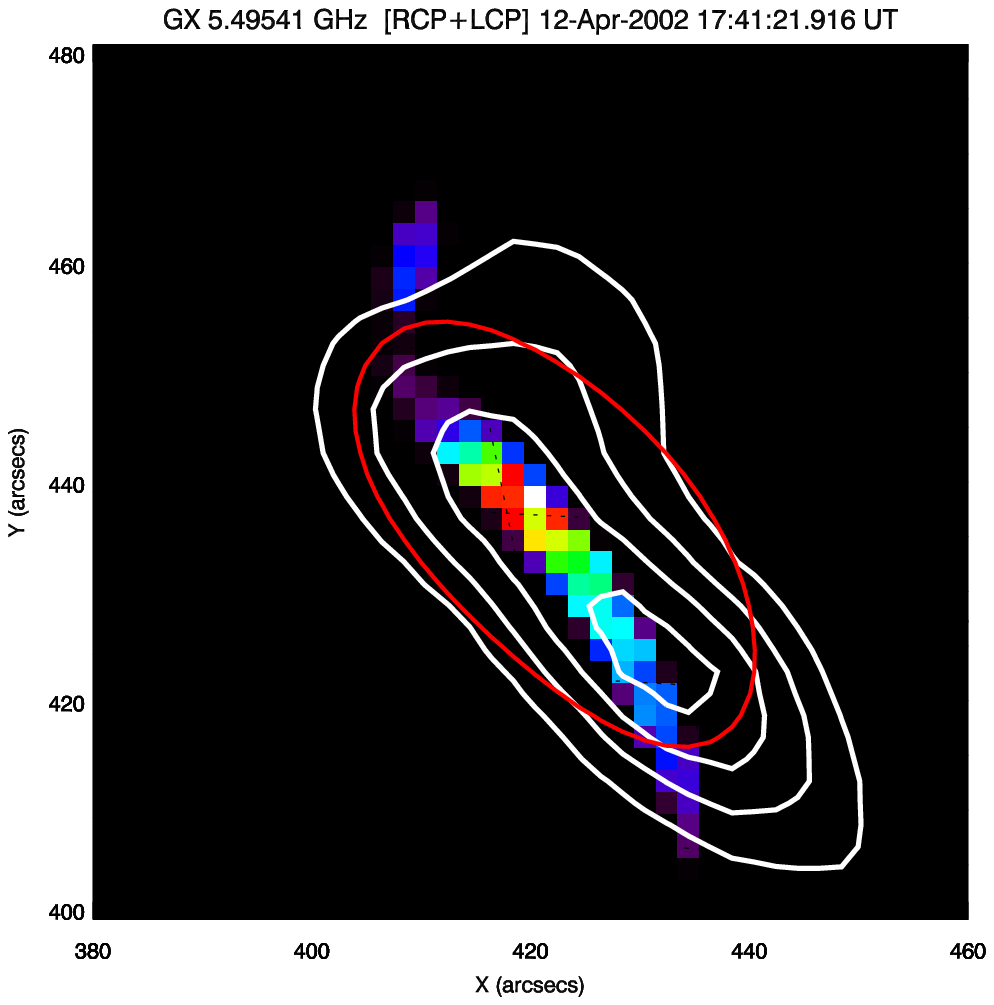}b
\includegraphics[width=0.31\textwidth,clip]{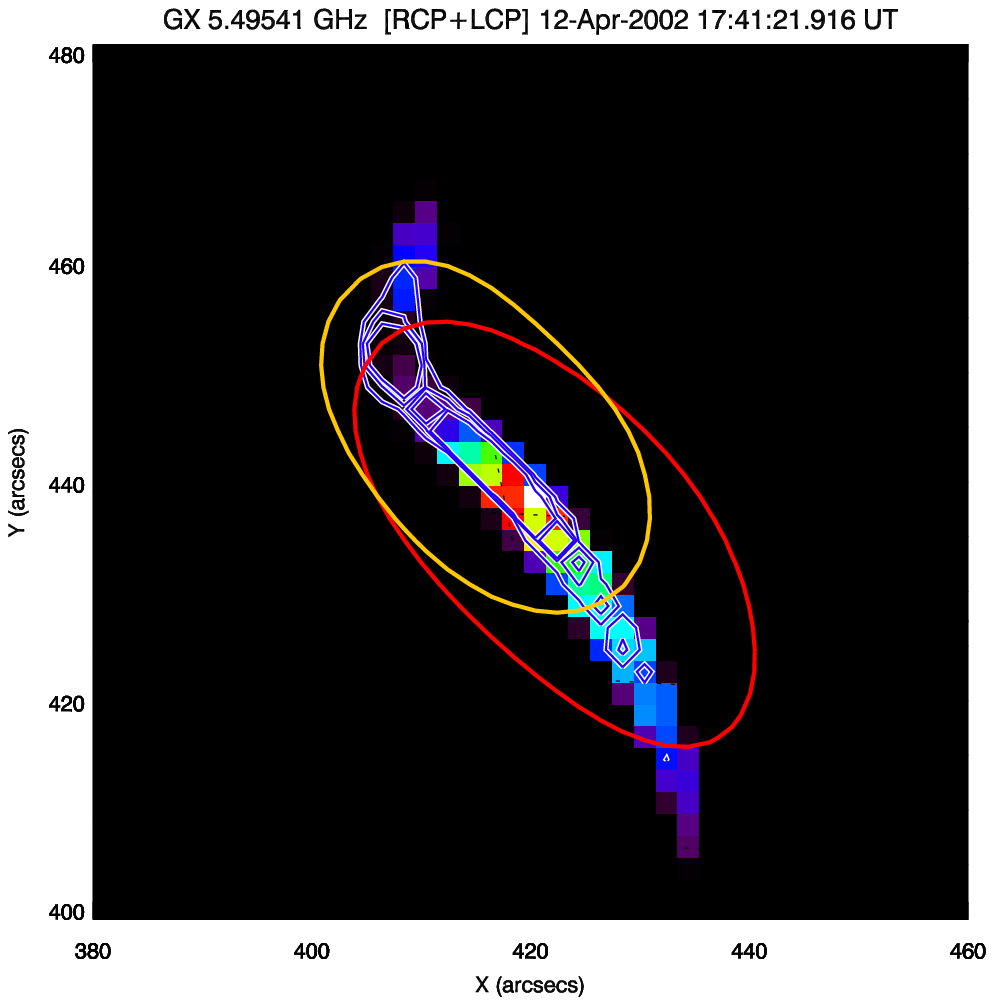}c\\
\includegraphics[width=0.3\textwidth,clip]{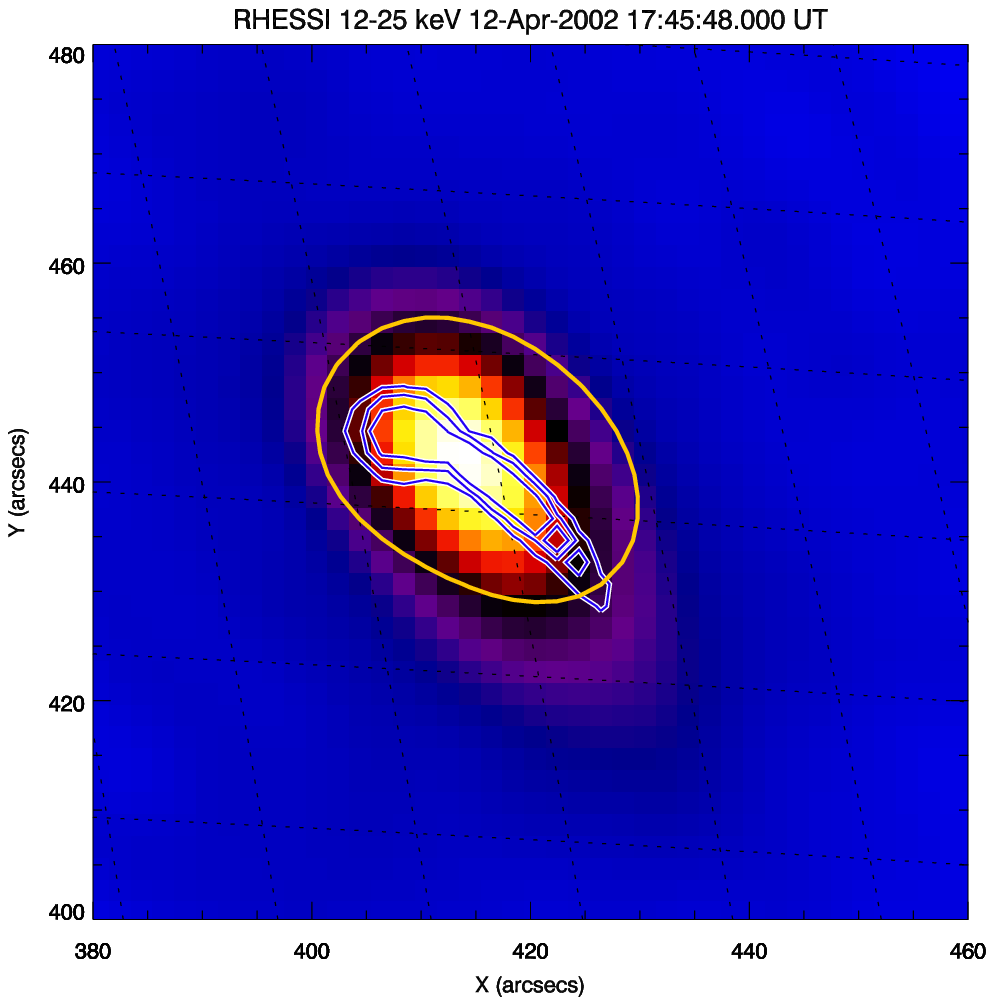}d
\includegraphics[width=0.315\textwidth,clip]{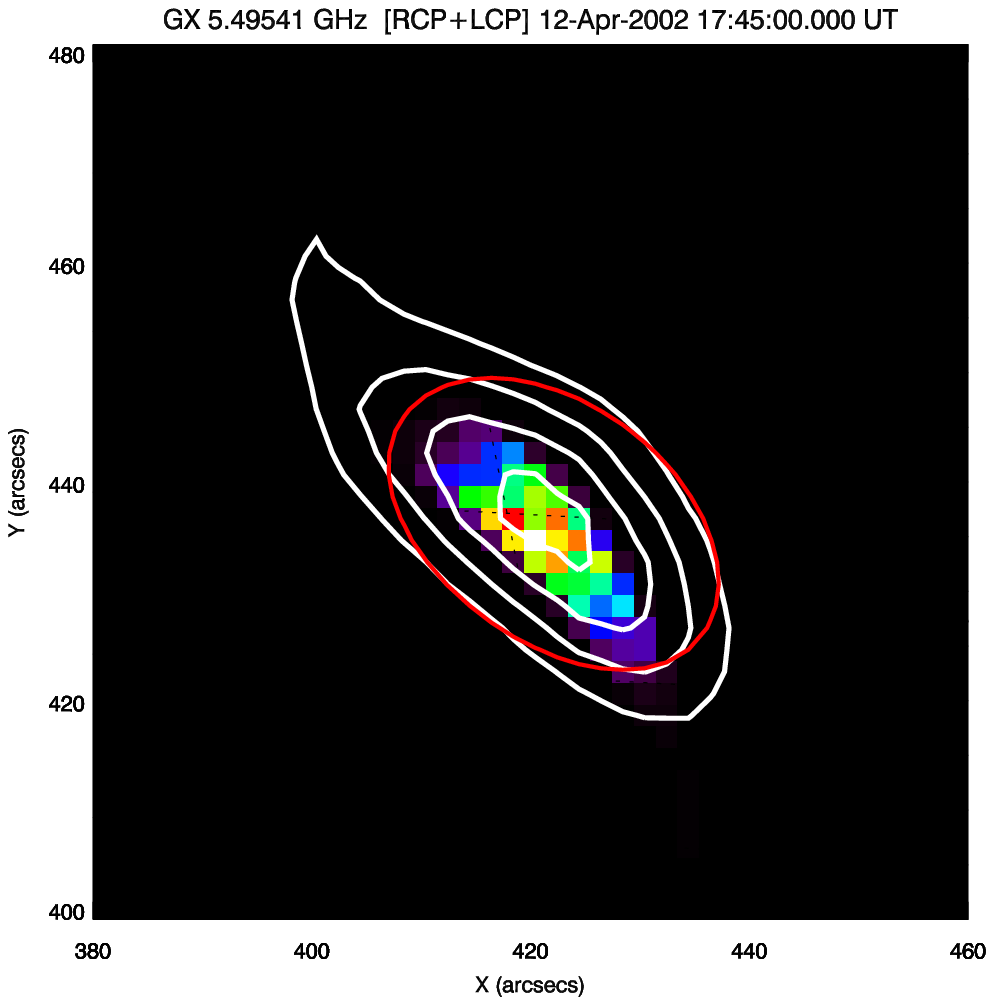}e
\includegraphics[width=0.31\textwidth,clip]{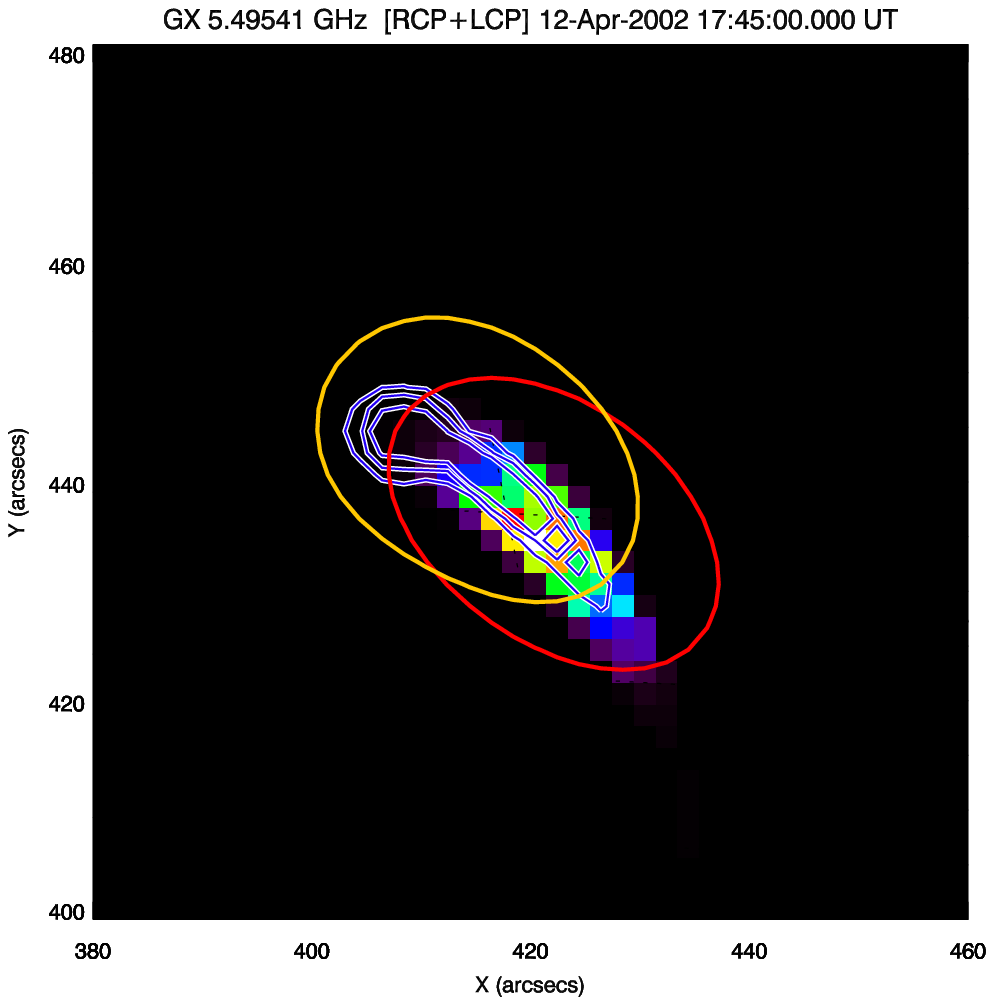}f
\caption{Comparison of the observed and synthetic images computed from the adopted 3D model {for the rise (a-c) and peak (d-f) phases}. a: [2,~7,~15]\% contours of the computed HXR emission at 15~keV (representative of the images at 12--25~keV; blue) on top of the \rhessi\ image (same as in Fig.~\ref{fig_images}); b: [30,~50,~70,~90]\% contours of the  OVSA time and frequency~2.8--5.6~GHz synthesis emission  (white; same as in Fig.~\ref{fig_images}) on top of the  computed \mw\ 5.6~GHz image; c: model HXR emission at 15~keV (blue) on top of the model \mw\ image at ~2.8--5.6~GHz. d: [5,~15,~30]\% contours of the computed HXR emission at 15~keV (representative of the images at 12--25~keV; blue) on top of the \rhessi\ image (same as in Fig.~\ref{fig_images}); e:  [30,~50,~70,~90]\% contours of the  OVSA~5.6~GHz emission  (white; same as in Fig.~\ref{fig_images}) on top of the  computed \mw\ 5.6~GHz image; f: model HXR emission at 15~keV (blue) on top of the model \mw\ image at ~5.6~GHz. Yellow and red contours show 30\% level of the model image convolved with the corresponding psf of \rhessi\ and OVSA, respectively. It is apparent that the modeled
\mw\ and HXR images computed from the same, \textit{energy independent}, spatial distribution of nonthermal electrons are \textit{displaced} relative to each other in exactly same sense as suggested by observations.
\label{fig_model_images_rise}
}
\end{figure*}

\begin{figure*} 
\includegraphics[width=0.99\textwidth,clip]{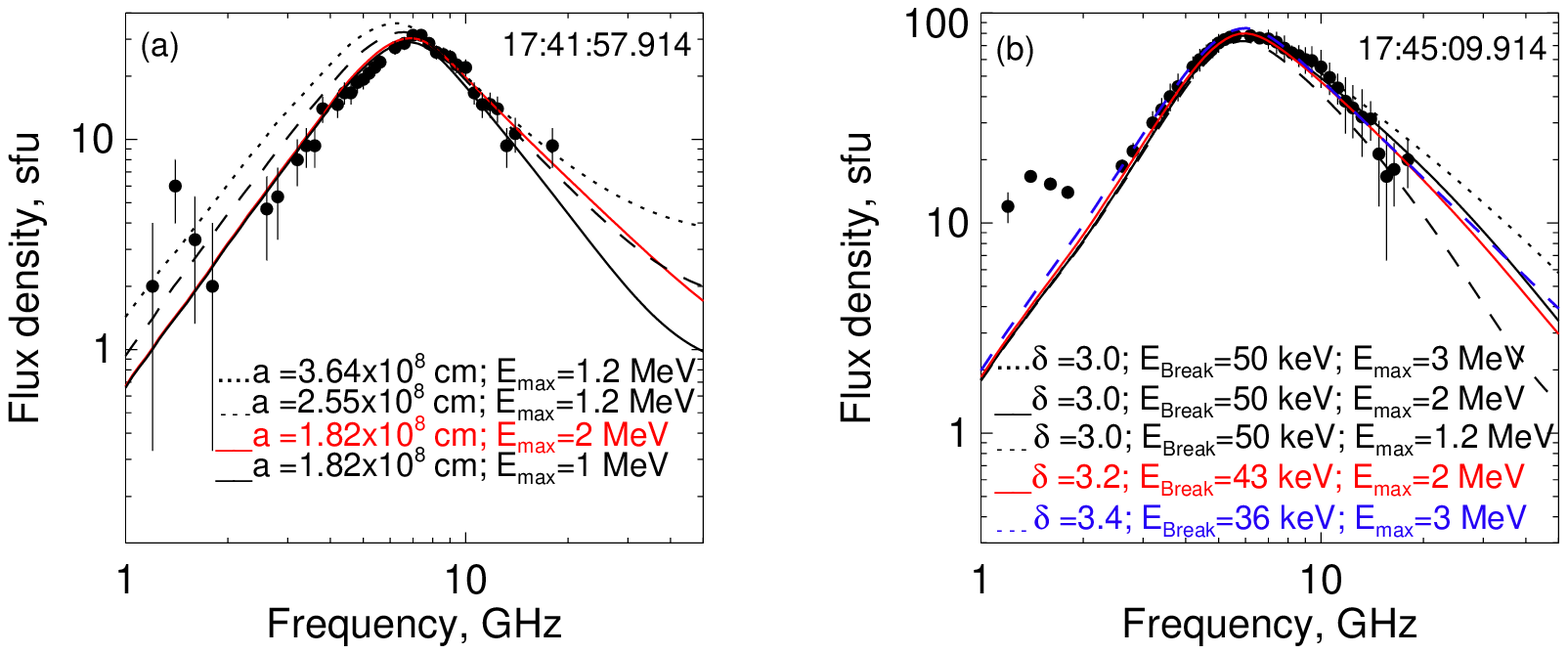} 
\caption{{(a):} Comparison of the \mw\ spectra obtained by  OVSA at the rise phase at 17:41:58~UT  and the image-integrated synthetic spectra computed from the adopted 3D model loop with $a\approx 1.82\cdot10^8$~cm,  {red} solid line{, and from the same 3D model but assuming a smaller values of the high-energy cut-off in the electron spectrum, $E_{\max} = 1$~MeV, black solid line}, and two models with thicker loops: $a\approx 2.55\cdot10^8$~cm (dashed line) and $a\approx 3.64\cdot10^8$~cm (dotted line; the model excess at the high frequencies is due to enhanced free-free emission from this thicker loop). The remarkable model-to-data agreement is apparent in case of the adopted model loop with $a\approx 1.82\cdot10^8$~cm. Thicker loops (with the same total number of the nonthermal electrons and the same $EM$) produce excessive low-frequency emission.  
{(b): a similar model-to-data comparison but for the peak phase of the event. The red solid curve is produced for the adopted model (Table~\ref{table:2}); other curves give some idea about uncertainties of the model parameters. {An additional decimetric component seen as a distinct spectrum enhancement at 1--2~GHz comes from a location displaced by roughly $20''$ SW from the main \mw\ source (see animation to Fig.~\ref{fig_images}); we do not have enough constraints to firmly interpret this extra component in our event, which can either be another GS source or be produced by a distinct emission process, such as plasma emission or transition radiation \citep{RTR}.}}
\label{fig_model_spectra_rise}
}
\end{figure*}

The needed input for GX Simulator is the instantaneous distribution of the nonthermal electron \textit{number density}, while from the thin-target spectral fit described above we obtained the \textit{electron flux}. To find the mean density of nonthermal electrons $\langle n_b\rangle=F_e/v$, {where $v$ is the velocity that corresponds to the low-energy cut-off in the X-ray fit, $E_0=18$~keV,} we note that for $n_0 V=2.25\times10^{37}$ specified above we get $F_e =0.50\times10^{55}/(n_0 V) \approx 2.2\times10^{17}$~el's~cm$^{-2}$~s$^{-1}$, thus for $v(18 keV)=0.78\cdot 10^{10}$~cm/s we obtain $\langle n_b\rangle=0.28\times10^{8}$~cm$^{-3}$ and the spectral index $\delta_{n1}= \delta_{l}+0.5 =7.2\pm0.5$, that is by 0.5 larger than the spectral index in Table~\ref{table:1}. To ensure this mean value over the model volume we adopted the peak number density (which is the immediate input for the GX Simulator) to be $n_b=10^{8}$~cm$^{-3}$; we checked by direct computation that the adopted model produces the synthetic X-ray spectrum consistent with the observed one. {However, we quickly find that} it is not possible to obtain a good fit to the \mw\ spectrum with {such a} steep electron energy spectrum, irrespective of other model assumptions. Unavoidably, we have to assume the existence of a spectral break-up above some energy, {to a lower} spectral index of about $\delta_{n2}\approx3.5$ to fit the high-frequency \mw\ spectral slope. {
The need for a flatter radio-derived electron spectrum relative to that derived from hard X-rays is common to many previous radio-to-X-ray comparisons
\citep[see, e.g.,][and references therein]{2000ApJ...545.1116S, 2011SSRv..159..225W}
}

The above consideration implies {that we try a} double power-law spectrum of the total number of nonthermal electrons with three parameters constrained so far---the total electron number and low-energy spectral index are fixed based on  the {described above X-ray fit to the} observed HXR emission, see Table~\ref{table:1}, while the high-energy spectral index is constrained by the high-frequency slope of the \mw\ spectrum.  Two other parameters, namely, the break energy and the maximum energy have not yet been defined. We do not have {a suitable constraint on the value of} $E_{\max}$, so we {choose a reasonably high value}, $E_{\max}\gtrsim1$~MeV. On the {other hand}, the break energy $E_{\rm break}$ {is constrained by the level of the high-frequency \mw\ emission. The adjustment of $E_{\rm break}$ also} has to be done consistently with matching the spectral peak frequency and flux level, which depend on the magnetic field, and with matching the low-frequency slope, which, in addition, depends on the source geometry---its size and nonuniformity.

It appears, however, not possible to achieve a reasonable match by adjusting the break energy in the nonthermal electron spectrum in the given magnetic loop.  When the high-energy electron numbers are boosted higher (lower $E_{\rm break}$) to fit the optically thin radio flux density, the model spectral peak frequency remains noticeably smaller than observed.  As is well known from the theory of GS radiation, a higher magnetic field strength in the model  is required to bring the peak frequency into a match with the observations.
{We seek such a modification of the model as follows:}
as noted earlier, we do not have vector magnetic field data on which to base the extrapolation, and we instead adjusted the $\alpha$ parameter to find a matching shape of the loop.  However, this does not necessarily provide the correct absolute field strength. Thus, there may be some freedom to further modify the selected magnetic field model. However, an arbitrary nonlinear transformation of the data cube will create a new data cube {that is no longer} a valid solution of the Maxwell equations, which is inappropriate for realistic modeling. The simplest allowable transformation of the magnetic data cube that maintains its validity is a linear scaling by a hopefully small constant factor. In our case we found that a factor of 3.3 
is sufficient to entirely remove the revealed discrepancy and obtain the correct \mw\ peak frequency.

Then, to obtain a correct \mw\ peak flux---to be specific, we consider {here} the \mw\ spectrum at the sub-peak at 17:41:58~UT---we have to adopt $E_{\rm break}=36$~keV. The last step is to use the low-frequency \mw\ slope to constrain the loop geometry. It appears that the low-frequency slope is relatively steep, which requires a relatively thin magnetic flux tube, with the reference radius $a\approx 1.82\cdot10^8$~cm; thicker loops overestimate the flux level at the low, optically thick, frequencies and  produce progressively broader spectrum and shallower low-frequency \mw\ spectrum slope in disagreement with observations.

At this point the 3D model of the flaring loop, illustrated in Fig.~\ref{fig_model_rise}, is fully specified (it is summarized in Table~\ref{table:2}), and allows computing the synthetic X-ray and \mw\ images and spectra. Fig.~\ref{fig_model_images_rise} displays the success of this model in reproducing the images: there is a very good agreement between locations and sizes of the modeled and observed images in both X-rays and \mw s, as well as the spatial displacement between the model X-ray and model \mw\ images, similar to the revealed observational relationship. This {nicely demonstrates that a} spatial displacement of the {sources can be} produced by the same spatial distribution of the nonthermal electrons at all energies; thus the {sole} reason for this displacement is {the} differing distribution of the ambient plasma density (controlling the X-ray emission) and the magnetic field strength (controlling the GS emission). Another remarkable success of the modeling is in reproducing the entire microwave spectrum---both optically thin and thick, as well as  the spectral peak itself; see Fig.~\ref{fig_model_spectra_rise}, left.

It is important to test how sensitive the model outcome is to variations of the input parameters. In particular, we found that we can get a comparably good \mw\ fit if we assume the emission measure and temperature to be two times bigger and smaller, respectively,  since for the obtained source sizes and magnetic field range the \mw\ emission is only weakly sensitive to the thermal plasma properties; thus, we do not obtain additional constraints on the parameters derived from the X-ray fit. However, the \mw\ data help to significantly constrain other parameters, unconstrained by the X-ray data. First,  it constrains the reference width of the fluxtube $a$, which specifies the transverse extent of the thermal plasma and nonthermal electrons. This is illustrated by Fig.~\ref{fig_model_spectra_rise}, left, which, in addition to the developed best fit model, also shows  spectra obtained from two thicker loops, having proportionally bigger effective source area. These thicker loops clearly overestimate the \mw\ spectrum at low frequencies and, thus, must be discarded. Another \mw-constrained ingredient is the nonthermal electron spectral shape at high energies, not accessible for the X-ray probing with either \rhessi\ or \kw\ data, which we illustrate below using emission at the flare peak phase. {These results nicely demonstrate the high degree of complementarity of} the X-ray and \mw\ data in constraining a 3D model of the flaring loop.

\subsection{3D modeling: the peak phase}

It is instructive to develop a similar model for the flare peak time {to evaluate what flare parameters change the most remarkably over the course of the flare}. Although the high-energy \rhessi\ emission is strongly contaminated by the radiation belt particles, we still can analyze the same low-energy interval as in the rise phase. It is interesting that both HXR and \mw\ images are displaced (towards each other) compared with their locations in the rise phase, which implies some change in the magnetic flux tube configuration and, thus, calls for {a change in the} LFFF extrapolated data cube. We found a required connectivity in the extrapolation with a slightly smaller {nominal force-free parameter, $\alpha\approx6.53\cdot10^{-10}$~cm$^{-1}$, than for the rise phase. Although the nominal $\alpha$ value turns to be smaller than for the rise phase model, the actual $\alpha$ value along the central field line of the flaring flux tube in now slightly larger, $(6.5\pm2.4)\times10^{-10}$, but this difference is not statistically significant.}  The overall steps needed to build the model and adjust the parameters to the point when the images and spectra are fit are the same as described above {for the rise phase model}; the overview of the model is given in Fig.~\ref{fig_model_rise}, e; the model is summarized in the last column of Table~\ref{table:2}.

Remarkably, it is again possible to obtain, in agreement with observations, the displaced HXR and \mw\ images (Fig.~\ref{fig_model_images_rise}) from the same spatial distribution of nonthermal electrons at all energies and, at the same time, have a good spectral fit from the model. Fig.~\ref{fig_model_spectra_rise}, right, illustrates to what extent the high-energy electron spectrum is constrained by the \mw\ data.  Although a rather good fit is obtained for our 'best-fit' model outlined in Table~\ref{table:2}, having $\delta_{n2}=3.2$, $E_{\rm break}=43$~keV, and $E_{\max} = 2$~MeV, the data are also consistent with a range of spectral indices between 3.0 and 3.4, and the break up and max energies shown in the plot. This gives a clear idea of the likely ranges of these parameters. {Note, however, that the model with $\delta_{n2}=3.4$, $E_{\rm break}=36$~keV, which gives the \mw\ spectrum perfectly consistent with the data shown in blue in Fig.~\ref{fig_model_spectra_rise}b, produces a noticeable excess of HXR emission around 30~keV; thus, such a model, even though it offers a nice fit to the \mw\ data, must nevertheless be  discarded based on the HXR data. }

\subsection{3D modeling: summary}

{The degree to which the \mw\ spectrum of the model matches the} observed spectra is {quite} remarkable. Indeed, the exact spectral shape (slope) in the optically thick part of the spectrum is controlled by the actual distribution of the magnetic field values and the viewing angles in the nonuniform radio source; it is not {at all guaranteed that it is possible} to match the low-frequency spectral slope even with realistic, sophisticated 3D modeling \citep[e.g.,][]{Nita_etal_2015, 2015SoPh..290...79K}. In the case presented here, the spectral fit is {excellent, which together with the spatial agreement in {\mw}s and HXRs strongly argues for the validity} of the entire 3D model including the magnetic flux tube and the thermal and nonthermal particle distributions.


\begin{table*}[ht]
\caption{Summary of the 3D model}
\begin{tabular}{l l l l}
\hline\hline
Parameter & Symbol, units &  Rise Phase &  Peak Phase\\ [0.5ex]
\hline
{\textit{Central field line}:} &  & \\
\quad Length      & $l$, cm  & $5.7\cdot10^9$ & $4.77\cdot10^9$  \\
\quad Force-free parameter: Nominal & 
                                        $\alpha/ (10^{-10}$cm$^{-1}$) &  $6.6$ &  $6.53$ \\
\quad \quad \quad \quad at the flux tube 
                                            & $\alpha/ (10^{-10}$cm$^{-1}$) &  $6.4\pm2.0$ &  $6.5\pm2.4$ \\
\quad Number of twists & $N_{twist}={\alpha l}/({4\pi})$ &  $0.29$ &  $0.25$ \\
\quad Magnetic field, FP1      & $B_{f-}(s=0)$,~G  & -382 & -112  \\
\quad Magnetic field, FP2      & $B_{f+}(s=l)$,~G  & 2726 & 1860  \\
\quad Magnetic field, LT      & $|B_{\rm ref}|$,~G  & 205 & 187  \\
\textit{Fluxtube:} &  &  \\
\quad Reference cross-section radius & $a=b$, cm & $1.82\cdot10^8$ & $5.8\cdot10^8$  \\
\quad Reference point location & $s_0$, cm & $-1.64\cdot10^9$ & $-5.23\cdot10^8$  \\
\quad Model volume; $\left[\int n_0 dV\right]^2/\int n_0^2 dV$ & $V$, cm$^3$ & $2.45\cdot10^{26}$ & $1.88\cdot10^{27}$  \\
{\textit{Thermal plasma}:} &  & \\
\quad Number density at central field line & $n_0$,  cm$^{-3}$ & $2.0\times10^{11}$ & $1.0\times10^{11}$  \\
\quad Mean number density; $\int n_0^2 dV/\int n_0 dV$ & $\langle n_0\rangle$,  cm$^{-3}$ & $9.2\times10^{10}$ & $3.7\times10^{10}$  \\
\quad Temperature      & $T$, MK  &  20 & 21  \\
\quad Parms of transverse distribution      & $p_0,~p_1,~p_2,~p_3$  &  2, 2, 0, 0 & 2, 2, 0, 0  \\
\textit{Nonthermal electrons:} &  &  \\
\quad Number density at central field line & $n_b$, cm$^{-3}$ &  $1.0\times10^{8}$ & $0.8\times10^{8}$   \\
\quad Mean number density; $\int n_b^2 dV/\int n_b dV$ & $\langle n_b\rangle$,  cm$^{-3}$ & $3.5\times10^{7}$ & $2.7\times10^{7}$  \\
\quad Parms of transverse distribution      & $p_0,~p_1,~p_2,~p_3$  &  2, 2, 0, 0 & 2, 2, 0, 0  \\
\quad Parms of distribution along the loop      & $q_0,~q_1,~q_2$  &  7, 0, 0 & 8, 0, -0.3  \\
\quad Total electron number & $N_b$, cm$^{-3}$ &  $3.66\times10^{33}$ & $0.99\times10^{34}$   \\
\quad Low-energy Cutoff & $E_0$, keV & 18  & 18   \\
\quad High-energy Cutoff  & $E_{\rm max}$, MeV & 2 & $2$  \\
\quad Break Energy & $E_{\rm break}$, keV & 36 & 43 \\
\quad Low-energy Index & $\delta_{n1}(<E_{\rm break})$ & 7.3 & 6.3 \\
\quad High-energy Index & $\delta_{n2}(>E_{\rm break})$ & 3.4 & 3.2 \\
[1ex]
\hline

\end{tabular}


\label{table:2}
\end{table*}

\section{Discussion}

The importance of the validated 3D flare model above is several fold.
A relatively simple, single flare loop model with a filling factor 1 is overall fully consistent with the {combined} X-ray, \mw, and context magnetic field data. Having the specified 3D model allows the study of how this same flare would look being observed from a different perspective; e.g., at the limb or disk center.
The overall combination of the X-ray and \mw\ data in this event can be quantitatively reproduced in the model that relies on the thermal and nonthermal electron parameters recovered from the spectral fit of the X-ray spectral data.

One of the model outcomes is the nonthermal electron spectrum break-up with flattening at high energies. We have mentioned that a similar flattening was reported earlier based on the \mw-to-X-ray slope comparisons \citep{2000ApJ...545.1116S, 2011SSRv..159..225W}. In the previous cases, however, the radio-derived and X-ray-derived spectral indices of nonthermal electrons were obtained using one or another approximation, e.g., from the thick-target fit of the footpoint X-ray emission or a simplified analytical Dulk-Marsh formula for the GS spectrum, which necessarily involves some level of uncertainty \citep[see, e.g.,][]{Kuznetsov_etal_2011}. Accordingly, in some of such cases the spectral index mismatches disappeared, when a more accurate treatment was used \citep{Fl_Kuzn_2010, Fl_etal_2011}.
In contrast, our modeling for the first time shows that the entire spectrum of the nonthermal electrons at the same spatial location is required to have a double power-law form with a precisely constrained break energy, as well as low- and high- energy spectral indices.

The spatial peak of the nonthermal electron distribution is significantly displaced from the location where the magnetic field is minimal along the loop spine: the latter is, in fact, very close to one of the loop footpoints, while the former is close to the geometrical center of the loop. This means that the spatial distribution of the accelerated electrons cannot be accounted for by purely magnetic trapping in the looptop, as this mechanism would imply the electron concentration in the region where the magnetic field is minimal. One possible interpretation for this behavior is that the nonthermal electrons are trapped by turbulence at the acceleration region of the flare. Note that exactly the same spatial distribution is adopted for the electrons at all energies---both 10s keV electrons producing the X-ray emission and hundreds keV to MeV electrons producing the \mw\ emission. However, this same spatial distribution, being convolved with either spatially-varying ambient plasma density (in the case of HXRs) or the magnetic field (in the case of microwaves), produce the spatially displaced X-ray and \mw\ images as observed.

It is interesting to compare the physical parameters of the model at the rise and peak phases, see Table~\ref{table:2}, to conclude which parameters change over the course of flare, and which parameters stay roughly the same, to better understand what drives this apparently single-loop flare. It appears that the central field line of the flaring loop {in the two stages evolves:} the rise phase is associated with a longer loop, having stronger magnetic field, and somewhat larger twist {({but slightly smaller} $\alpha$)} than the loop associated with the peak phase. Then, the fluxtube at the peak phase is noticeably thicker than the fluxtube at the rise phase, with correspondingly bigger model volume. Thus, we see a significant variation of the source geometry. On the contrary, {many of the} plasma and nonthermal electron parameters remain almost unchanged between the rise and peak phases, {such as their} spatial distributions, their temperature, and number density. This implies that both \mw\ and HXR fluxes from the flare go up primarily because of the flare volume increase, rather than to the rise of the radiating particle densities. This further implies that the energy release and particle acceleration processes are spreading within the involved volume due, perhaps, to a reconnection spreading process. Although the number density of the accelerated electrons remains roughly the same, their spectrum evolves towards a more energetic, harder distribution: both spectral indices, $\delta_{n1}$ and  $\delta_{n2}$, become smaller at the peak phase.
These findings are in line with the conclusion made earlier that such coronal HXR sources originate from the coronal thick-target regions and represent the very sites of the electron acceleration in flares.

\acknowledgments

This work was supported in part by NSF grants  AGS-1250374, AGS-1262772, AGS-1348513, and AGS-1408703 and NASA grants  NNX13AG13G and NNX14AC87G to New Jersey
Institute of Technology. We are grateful to Dr. Valentin Pal'shin for providing the \kw\ data for this event available due to RFBR  grants 15-02-03835 \& 15-02-03717 and to Dr. Eduard Kontar for highly valuable discussions. 

\bibliographystyle{apj}
\bibliography{20020412_flare_ref,fleishman}

\end{document}